\documentclass[onecolumn,showpacs,preprintnumbers,amsmath,amssymb,bbold]{revtex4}
\usepackage{epsfig,amsopn}
\usepackage{graphicx}
\usepackage{epstopdf}
\usepackage{sidecap}
\usepackage{amsmath,amssymb}
\usepackage{amsthm}
\usepackage{enumerate}
\usepackage{bbold}
\usepackage{color}

\newcommand\bea{\begin{eqnarray}}
\newcommand\eea{\end{eqnarray}}
\newcommand\beq{\begin{equation}}  
\newcommand\eeq{\end{equation}}

\begin{document}

\title{Slave fermion formalism for  the tetrahedral spin chain }
\author{Priyanka Mohan and Sumathi Rao}
\affiliation{Harish-Chandra Research Institute, Chhatnag Road, 
Jhusi, Allahabad 211 019, India }

 \begin{abstract}
 
We use the SU(2) slave fermion approach to study a tetrahedral spin 1/2 chain,  which
is a one-dimensional generalization of the two dimensional Kitaev honeycomb model.  
Using the mean field theory, coupled with a gauge fixing procedure 
to implement the single occupancy constraint, we obtain the phase diagram of the model.
We  then show that  it matches the exact results obtained earlier using the Majorana fermion representation.
We also compute the spin-spin correlation in the gapless phase  and show that it is a spin liquid.
Finally, we  map the one-dimensional model in terms of the slave fermions to the model of 1D $p$-wave superconducting
model with complex parameters and  show that the parameters of our model fall in the topological trivial regime and hence
does not have edge Majorana modes.

 \end{abstract} 

\maketitle

\section{Introduction}

The study of quantum spin liquids in the context of frustrated magnetism in many body theory has attracted attention for
many years~\cite{reviews}. The impetus for the work in this area came from the understanding that these are examples where the localised
moments  have  strong correlations and lead to remarkable collective  phenomena such as emergent gauge fields
and fractional excitations. There exists well developed theoretical tools such as gauge theories~\cite{gauge}, slave boson~\cite{sb} and
slave fermion~\cite{sf} methods,  quantum dimer models~\cite{qd} and various numerical methods~\cite{numerics} to study such models.  

 Current
interest in this field has been enhanced due to the discovery of a remarkable exactly solvable spin 1/2 model  in two
dimensions known as as the Kitaev model with a gapless $Z_2$ spin liquid phase \cite{2dkitaev} with fractionalised excitations
(anyons and non-abelian anyons).
The ground state has degeneracy and  topological order which makes the model relevant for quantum computation~\cite{qc}.  Moreover, it
is an excellent model to study features of the spin liquid ground state of the gapless phase.
The disadvantage of the exactly solvable model is that it does not give a systematic way of dealing
with models which are not exactly solvable. Hence, it is of use to understand the connection between
the earlier methods of studying spin liquid, such as using spin waves, slave fermions, etc and the exact
solutions. Motivated by this, Burnell and Nayak~\cite{bn} studied the Kitaev model in two dimensions using the SU(2) slave
fermion formalism. They were able to reproduce the results and understand  the physics of the model  using the slave
fermion band structure. Around  the same time, Mandal et al.~\cite{mandal} related the
Kitaev model to the SU(2) gauge theory of the RVB model and showed how the residual $Z_2$ invariance
was related to Kitaev's Majorana representation of the spins.

On the other hand, there has also been a lot of interest in other  generalizations of the Kitaev model on other lattices,
and in other dimensions, as long as they have coordination number three~\cite{genKitaev,saket1,saket2}. In fact, besides being of theoretical interest,
such models can be realised in cold atom systems~\cite{coldatoms} and quantum circuits~\cite{qcircuits}.
Most of these models have been studied mainly by using some variant of the Majorana representation of the
spins to get some exact results.

Spin models  in one dimension also have a long history of having  been studied by various methods~\cite{onedmodels}. 
It is, hence, of interest to understand the relation between various  methods. In this paper, we focus on the Kitaev tetrahedral model,
earlier studied in terms of the Majorana representation of spins by Saket et al~\cite{saket1}.
Their aim was to see whether Majorana modes
could be created and manipulated by tuning the local fluxes and parameters of the theory and for this,
they represented  the Hamiltonian in terms of Majorana fermions using the Jordan-Wigner transformation. 
Our main aim in this paper is in examining how the phase diagram of the Kitaev tetrahedral model  in one dimension,
can be obtained through more conventional  methods.
For this,  we use a different representation for the spins in terms of the slave fermions (usually called spinons)
in terms of which the Hamiltonian becomes four fermion-like. This  model is equivalent to the spin model
only in the singly occupied sector (each site is occupied by a single electron), and this constraint is enforced as
an $SU(2)$ Gauss law constraint. Then using 
the Hubbard-Stratanovich decoupling in the hopping and superconducting channels, 
we  obtain  the phase diagram of the model and find that the gapless phase is confined to circles on a cone,
with all other regions being gapped. This agrees with the earlier results obtained by writing the spins in terms
of Majorana operators~\cite{saket1}
Finally, we map the tetrahedral model to an effective 1 D $p$-wave superconducting model (the `other'  or the one-dimensional
Kitaev model~\cite{onedkitaev}), which is the prototype model giving rise to Majorana modes at the edges when the chemical potential is
tuned to the topological regime.
 We then show that the parameters of the tetrahedral model
implies that the ground state of the effective one-dimensional Kitaev-like model is  in the topologically trivial phase without  Majorana modes at the edges.

\section{Model}

\begin{center}
   \begin{figure}[!ht]
\includegraphics[width=15cm]{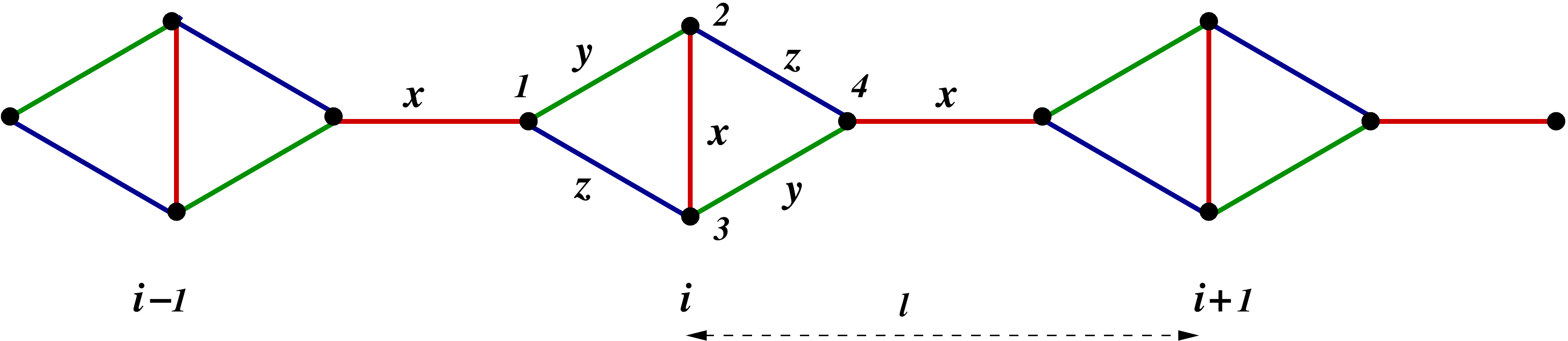}
\caption{The tetrahedral chain with four sites per unit cell. ${\bf x}$ denotes the quadratic spin bond $\sigma^x\sigma^x$ with the spins at
the two adjacent sites and similarly  ${\bf y}$ and ${\bf z}$ denote  $\sigma^y\sigma^y$ and  $\sigma^z\sigma^z$ respectively. }
 \label{Fig:Tetra_chain}
\end{figure}
 \end{center}

Kitaev's honeycomb model can be generalized to a variety of lattices which have coordination number three.
Here, we study the tetrahedral chain  introduced in Ref.[\onlinecite{saket1}], made by connecting a series of
plaquettes as shown  in Fig.\ref{Fig:Tetra_chain}.  Each plaquette or unit cell has four sites and the $x,y$ and $z$ bonds
are indicated in the figure. 
The Hamiltonian for such a model is,
 \begin{equation}
 H=-\sum_i [J_x(\sigma^x_{i-1,4}\;\sigma^x_{i,1}+\sigma^x_{i,2}\;\sigma^x_{i,3})
 +J_y(\sigma^y_{i,1}\;\sigma^y_{i,2}+\sigma^y_{i,3}\;\sigma^y_{i,4})
 +J_z(\sigma^z_{i,1}\;\sigma^z_{i,3}+\sigma^z_{i,2}\;\sigma^z_{i,4})].
  \label{Eqn:tetra_H}
 \end{equation}
 where the index $i$ denotes the site of the plaquette in the chain  and the second index (1...4) denotes the lattice site on  the plaquette or the unit cell.
The spin operators $\sigma^{x,y,z}_i$ can be represented using fermion operators using the relation, $\hat{\sigma}^\kappa_i=\frac{1}{2}f^{\dagger}_{i\alpha}
 \sigma^\kappa_{\alpha \beta} f_{i\beta}$. Here, $\sigma$ denotes the spin operators on the LHS and the Pauli matrices on the RHS. The 
 operators $f_{i\alpha}$ and $f^{\dagger}_{i\alpha}$ are the fermion annihilation and creation operators. The index $i$ denotes
 the site, indices $\kappa=(x,y,z)$  denote the three components of the spin and the indices $\alpha,\,\beta = \uparrow,\downarrow$ denote
 the spin projections. The model was introduced as a toy model to understand the manipulation of Majorana fermions, so it
 was studied analogous to the Kitaev model by representing the spins in terms of Majorana fermions. 
 The symmetries of this model were analysed and it was shown that every state of the system had a degeneracy of $2^{N}$,
 where $N$ is the number of unit cells.
Solutions for the Majorana edge states that occur in this model were presented and the creation  and manipulation of Majorana modes
were discussed~\cite{saket1}. 
 
 However, our focus in this paper is different.  We wish to analyse the model as a spin model and understand its phases, following the work of Burnell and Nayak~\cite{bn}, 
 who analysed the two dimensional Kitaev model in terms of slave fermions. So our
 starting point is to use the slave fermion formalism where the spins are written in terms of standard fermions, rather than Majorana fermions, $i.e.$,
 we rewrite the quadratic terms of the Hamiltonian in Eq.\ref{Eqn:tetra_H}, the
 $\sigma^x_i \sigma^x_j$, $\sigma^y_i \sigma^y_j$ and $\sigma^z_i \sigma^z_j$,  on the $x,y$ and $z$ links respectively,
 in terms of standard  fermion operators.  The main point to note here is that unlike the Heisenberg model where there
 is a rotational spin symmetry at each link, here the rotational symmetry is broken and the spin along the three links at each site
 is different.  For the $\sigma^x_i \sigma^x_j$, we find 
 \beq
 \sigma^x_i \sigma^x_j = -\frac{1}{4}[f^{\dagger}_{i\uparrow}f^{\dagger}_{j\uparrow}f_{i\downarrow}f_{j\downarrow}+
 f^{\dagger}_{i\downarrow}f^{\dagger}_{j\downarrow}f_{i\uparrow}f_{j\uparrow}+
 f^{\dagger}_{i\uparrow}f_{j\uparrow}f^{\dagger}_{j\downarrow}f_{i\downarrow}+
 f^{\dagger}_{i\downarrow}f_{j\downarrow}f^{\dagger}_{j\uparrow}f_{i\uparrow}]~.
 \eeq
 Similar terms can be obtained for $\sigma^y_i \sigma^y_j$ and $\sigma^z_i \sigma^z_j$ as shown in the appendix.
 Note that there is a redundancy in representing the spins in terms of fermions. The fermionic operators
 reproduce the Hilbert space of the spin operators  
 only when  double occupancy and no occupancy are projected out  i.e., only
 in the singly occupied subspace.
Hence, this form is not unique and we have to satisfy the conditions, 
 $n_{i\uparrow}+n_i{\downarrow} =1$, $f_{i\uparrow}f_{i\downarrow}=0=f^\dagger_{i\uparrow}f^\dagger_{i\downarrow}$ on each site~\cite{bn}.
 The choice of a particular form can  also be understood as a gauge fixing condition, since the subspace of the single occupancy and
 the subspace of the no occupancy and double occupancy can be shown to form two independent representations of $SU(2)$~\cite{mandal}.

\section{Mean field Hamiltonian and phase diagram}

We now perform a mean field analysis of the Hamiltonian by introducing the Hubbard-Stratanovich ($\mathcal{HS}$)  fields in the 
hopping and the $p$-wave superconducting channels. As explained in Ref.[\onlinecite{bn}], in the
absence of spin-rotational symmetry, independent $\mathcal{HS}$ fields have to be introduced for the fermion operators
on the $x,\;y$ and $z$ links. The $\mathcal{HS}$ transformed form of the spin operators 
in terms of four   bosonic $\mathcal{HS}$ fields ($\phi_i$ and the $\theta_i$ fields) evaluated on the $x$ links in the hopping
and superconducting channels is given explicitly in  Appendix A.    
In the mean field limit, the $\mathcal{HS}$ fields take their expectation values. More explicitly, we parametrize the expectation values by the following 
hopping parameters and superconducting order parameters, similar to the 2D model:
$\langle \phi_1+i\phi_2\rangle= t^x_\uparrow$, 
$\langle \phi_1-i\phi_2\rangle= t^x_\downarrow$, $\langle \theta_1+i\theta_2\rangle= \Delta^x_\uparrow$,
$\langle \theta_1-i\theta_2\rangle= \Delta^x_\downarrow$.  
Similar terms can be written for    $J_y \sigma^y_i \sigma^y_j$ and $J_z\sigma^z_i \sigma^z_j$ terms.

Using these expressions, the original Hamiltonian in Eq.\ref{Eqn:tetra_H} can be written as
\begin{eqnarray}
 H_x &=& J_x(\sigma^x_{i-1,4}\;\sigma^x_{i,1}+\sigma^x_{i,2}\;\sigma^x_{i,3})=-\frac{16}{J_x}(|\phi_1|^2+|\phi_2|^2+|\theta_1|^2+|\theta_2|^2)\nonumber\\
 &+&\sum_{\alpha} [t^x_\alpha \;(f^{\dagger}_{i-1,4\;\alpha}f_{i,1\;\alpha}+f^{\dagger}_{i,2\;\alpha}f_{i,3\;\alpha}) +
\Delta^x_\alpha\; (f^{\dagger}_{i-1,4\;\alpha}f^{\dagger}_{i,1\;\alpha}
 +f^{\dagger}_{i,2\;\alpha}f^{\dagger}_{i,3\;\alpha}+c.c.] \nonumber \\
  \label{Eqn:Jx_HS_2}
  \end{eqnarray}
with similar terms for $J_y \sigma^y_i \sigma^y_j$ and $J_z\sigma^z_i \sigma^z_j$.
Assuming periodic boundary conditions and after Fourier transforming in the following way, 
\begin{equation}
 f_{i,j\;\alpha}=\frac{1}{\sqrt{N}}\sum_q e^{iq r_i}  f_{q,j\;\alpha}, \qquad f^{\dagger}_{i,j\;\alpha}=\frac{1}{\sqrt{N}}\sum_q e^{-iq r_i}  f^{\dagger}_{q,j\;\alpha}
 \label{Eqn:FT}
\end{equation}
we get the final Hamiltonian $H$.
Since the terms with spin indices $\uparrow$ and $\downarrow$ decouple in the Hamiltonian, the total Hamiltonian,
$H$  in Eq.\ref{Eqn:tetra_H} can be written as $H_0+ \sum_q(\psi^\dagger_{q \uparrow}H_{\uparrow}\psi_{q \uparrow}+
\psi^\dagger_{q \downarrow}H_{\downarrow}\psi_{q \downarrow})$, where $H_0 $  is the Gaussian term of the $\mathcal{HS}$ 
fields and  the independent Hamiltonians
for the $\uparrow$ and $\downarrow$ spins are given by
\begin{equation}
H_\alpha =\frac{1}{2} \sum_q
\begin{pmatrix}
  0 & t^y_\alpha & t^z_\alpha & t^{x^\ast}_\alpha e^{-iql} & 0 & \Delta^y_\alpha & \Delta^z_\alpha & -\Delta^x_\alpha e^{-iql} \\
 t^{y^\ast}_\alpha & 0 & t^x_\alpha & t^z_\alpha & -\Delta^y_\alpha & 0 & \Delta^x_\alpha & \Delta^z_\alpha \\
 t^{z^\ast}_\alpha & t^{x^\ast}_\alpha & 0 & t^y_\alpha & -\Delta^z_\alpha & -\Delta^x_\alpha & 0 & \Delta^y_\alpha \\
 t^x_\alpha\;e^{iql} & t^{z^\ast}_\alpha & t^{y^\ast}_\alpha & 0 & \Delta^x_\alpha e^{iql} & -\Delta^z_\alpha & -\Delta^y_\alpha & 0 \\
 0 & \Delta^{y^\ast}_\alpha & -\Delta^{z^\ast}_\alpha & \Delta^{x^*}_\alpha e^{-iql} & 0 & -t^{y^\ast}_\alpha & -t^{z^\ast}_\alpha & -t^x_\alpha e^{-iql} \\
 -\Delta^{y^\ast}_\alpha & 0 & -\Delta^{x^\ast}_\alpha & -\Delta^{z^\ast}_\alpha & -t^y_\alpha & 0 & -t^{x^\ast}_\alpha & -t^{z^\ast}_\alpha \\
 \Delta^{z^\ast}_\alpha & \Delta^{x^\ast}_\alpha & 0 & \Delta^{y^\ast}_\alpha & -t^z_\alpha & -t^x_\alpha & 0 & -t^{y^\ast}_\alpha \\
 -\Delta^{x^\ast}_\alpha e^{iql} & \Delta^{z^\ast}_\alpha & -\Delta^{y^\ast}_\alpha & 0 & -t^{x^\ast}_\alpha e^{iql} & -t^z_\alpha & -t^y_\alpha & 0 \\
\end{pmatrix}~.
\label{Eqn:H_matrix}
\end{equation}
Here $\alpha$ is the spin index and can be $\uparrow$ or $\downarrow$. Note that the Hamiltonian is twice as large,  $8\times8$ instead of
$4\times 4$,  as it was in the case of the two dimensional model~\cite{bn}.  This is due to the fact that in the one dimensional model,
there are both intra-plaquette bonds and inter-plaquette bonds for $J_x\sigma^x\sigma^x$ terms, which have different Fourier transforms.
The $\psi_{q,\alpha}$ and $\psi^{\dagger}_{q, \alpha}$ have the forms,
\begin{eqnarray}
 \psi_{q\;\alpha} &=& \begin{pmatrix}
f_{q,1\;\alpha}& f_{q,2\;\alpha}& f_{q,3\;\alpha} f_{q,4\;\alpha}&
f^\dagger_{-q,1\;\alpha}& f^\dagger_{-q,2\;\alpha} &f^\dagger_{-q,3\;\alpha}& f^\dagger_{-q,4\;\alpha}
\end{pmatrix}^T \nonumber \\ {\rm and} \quad
 \psi^\dagger_{q\;\alpha} &= &\begin{pmatrix}
f^{\dagger}_{q,1\;\alpha}&f^{\dagger}_{q,2\;\alpha}&f^{\dagger}_{q,3\;\alpha}&f^{\dagger}_{q,4\;\alpha}
&f_{-q,1\;\alpha}&f_{-q,2\;\alpha}&f_{-q,3\;\alpha}&f_{-q,4\;\alpha}\\
\end{pmatrix}
\label{Eqn:H_psidagger}
\end{eqnarray}
The factor of 1/2 in the first line is because the expression counts each term in the Hamiltonian twice.

Note also that we have not yet imposed the constraint of single occupancy. Enforcing the single occupancy constraint, $n_{i\;\uparrow}+n_{i\;\downarrow}=1$
on $z$ bonds is much easier than $x$ and $y$ bonds. This is due to the fact that fermionization of $\sigma^z_i$ gives
$ \frac{1}{2}(n_{i\;\uparrow}-n_{i\;\downarrow})$. The rest of the Hamiltonian has to be projected to a singly occupied subspace. One of the ways to impose  the 
conditions is by using Lagrange multipliers. The form of the Lagrange multipliers is  then calculated self-consistently. 
However, as was shown by Mandal, Shankar and Baskaran [\onlinecite{mandal}],  the model has an
$SU(2)$ gauge symmetry and the single occupancy constraint is equivalent to imposing the $SU(2)$ Gauss law.
In fact, Kitaev\rq{}s  model on the 2 dimensional honeycomb lattice is left with static $Z_2$ gauge fields even after
imposing the single occupancy constraint. So the Kitaev model can also be thought of as an SU(2) gauge theory,
where the spin operators fix the gauge, but still leave an unbroken $Z_2$ symmetry.

The next step is to find the self-consistent mean field solution. 
At this stage, our model has twelve   bosonic fields  - $t_{\uparrow/\downarrow}^\kappa, \Delta_{\uparrow/\downarrow}^\kappa$ where $\kappa =x,y,z$ (four  on  each link corresponding
to hopping $t$ of the $\uparrow$ or $\downarrow$ electron or corresponding to the superconducting
order parameter  $\Delta_{\uparrow/\downarrow}$). 
We note that since the pairing
is $p$-wave, the expectation values of the pairing potential within the plaquette is zero (i.e. $\Delta_{k,k^\prime}$ for $k$=$k^\prime$ in $\Delta_{k,k^\prime} c^\dagger_k c^\dagger_{k^\prime}$ ). Hence, we can set $\Delta^y_\alpha=\Delta^z_\alpha=0$ everywhere. But $\Delta^x_\alpha=0$ only for 
sublattice indices $2,3$, which are within a plaquette. It is non-zero when the sublattice indices are $1,4$ and we will denote this now simply by $\Delta^x_\alpha$ without
the index $1,4$. 

We now only have the    bosonic fields  $t_{\alpha}^\kappa$ where $\kappa =x,y,z$ 
and the single $\Delta_\alpha$ field ($\alpha$ for each spin) corresponding to the superconducting
order parameter.  We can write down the saddle point equations for these
fields in terms of the fermions $f_{l,\alpha}, f^\dagger_{j,\alpha}$ as
\beq
 t_{\alpha}^\kappa= \frac{J_\kappa}{4}\langle f^\dagger_{j,\alpha} f_{l,\alpha} \rangle, 
\quad \Delta_{\alpha}^\kappa= \frac{J_\kappa}{4}\langle f_{j,\alpha} f_{l,\alpha}\rangle
\eeq
where   $(l$-$j)=(i,2)$-$(i,3) $ and $(i-1,4)$-$(i,1)$ for the $x$- bonds, $(l$-$j)=(i,1)$-$(i,2) $ and $(i,3)$-$(i,4)$ for the $y$-bonds  and 
 $(l$-$j)=(i,2)$-$(i,4) $ and $(i,1)$-$(i,3)$ for the $z$- bonds.
Now, to satisfy these equations, there are various possible choices that we can make.  For the ground state however,
we would expect maximal symmetry and minimum energy.  Guided by this, we find that we can make the following choices - 
 \begin{eqnarray}
 &&   t^x_{\downarrow}=i\frac{J_x}{4},\quad \Delta^x_\downarrow=-i\frac{J_x}{4} \quad \text{for $(i-1,4)$-$(i,1)$ bond.}  \nonumber\\
 && t^x_\uparrow = 0,\quad \Delta^x_\uparrow =0 \quad \text{for $(i-1,4)$-$(i,1)$ bond.}  \nonumber\\
&&   t^x_{\alpha}=i\frac{J_x}{2},\quad \Delta^x_\alpha=0 \quad \text{for $(i,2)$-$(i,3)$ bond.}
\eea
  for the $x$-fields.  Here, we have differentiated between the intra-cell links and the inter-cell links. But for the $y$ and
 $z$ bonds, there is no such difference and we can simply choose
 \bea
&&   t^y_{\alpha}=i\frac{J_y}{2},\quad \Delta^y=0 \nonumber\\
&&   t^z_{\alpha}=-i\frac{J_z}{2},\quad \Delta^z=0 ~.
\label{meanfield}
   \end{eqnarray}
This choice is similar to the choice made by Burnell and Nayak\cite{bn} and reproduces the
exact result in terms of the Majorana fermions.  However, it is not a unique choice.  
In terms of the original lattice model, there are two triangular plaquettes in each 
unit cell and a conserved $Z_2$ flux associated with each plaquette, with the fluxes being given by
\beq
W_i^L = \sigma_{i,1}^x \sigma_{i,2}^y \sigma_{i,3}^z, \quad W_i^R = \sigma_{i,4}^x \sigma_{i,3}^z \sigma_{i,2}^y.
\eeq
However, the single particle spectrum only depends on the product of the fluxes $W_i^L\times W_i^R$. 
So for instance, we can choose $W_i^R = \pm1$ for all $i$ and $W_i^L =\pm 1$ for all $i$. Both the cases
$W_i^L=1=W_i^R$ and $W_i^L= -1=W_i^R$ will give the same result giving rise to a degeneracy of two for each plaquette.
This  of course, leads to an infinite degeneracy in the thermodynamic limit. However, if we only count configurations with a
definite value of the product $W_i^L\times W_i^R$, there still exists a  degeneracy in the thermodynamic limit.
In other words, besides  the indistinguishability of the individual fluxes
in the product of fluxes at each plaquette, there is a  gauge symmetry left in the model
after enforcing the single  occupancy constraint. This is the loop operator which can wind around the one-dimensional lattice 
(with periodic boundary conditions)  defined by
\beq
W_{\rm loop(z)}= \sigma_1^z \sigma_2^z\sigma_3^z \dots
\eeq
where the loop is defined by going along the  bonds with $J_x\sigma^x\sigma^x$ and $J_y\sigma^y\sigma^y$.
This operator commutes with the Hamiltonian and represents a conserved quantity, and can be +1 or -1,
giving a degeneracy of 2. Note that we could have defined other loop operators
\beq
W_{\rm loop(xyz)}= \sigma_1^z \sigma_2^x\sigma_3^y \dots
\eeq
where the loop is defined by going along the bonds with $J_x\sigma_x\sigma_x$, $J_y\sigma_y\sigma_y$ and  $J_z\sigma_z\sigma_z$ or
\beq
W_{\rm loop(y)}= \sigma_1^y \sigma_2^y\sigma_3^y \dots
\eeq
where the loop is defined by going along the bonds with $J_x\sigma^x\sigma^x$ and $J_z\sigma^z\sigma^z$.
But these do not result in independent conserved operators, because they can be obtained by multiplying
$W_{\rm loop(z)}$ with suitable combinations of $W_i^L W_i^R$. So the degeneracy in the thermodynamic limit of the model
(with the caveat mentioned above) is two.

The final Hamiltonian is thus given by
      \begin{equation}
H_\downarrow =\frac{i}{4}
\begin{pmatrix}
  0 & J_y & -J_z & -\frac{J_x}{2}\;e^{-iql} & 0 & 0 & 0 & \frac{J_x}{2}\;e^{-iql} \\
 -J_y & 0 & J_x & J_z & 0 & 0 & 0 & 0 \\
 J_z & -J_x & 0 & J_y & 0 & 0 & 0 & 0 \\
 \frac{J_x}{2}\;e^{iql} & -J_z & -J_y & 0 & -\frac{J_x}{2}\;e^{iql} & 0 & 0 & 0 \\
 0 & 0 & 0 & \frac{J_x}{2}\;e^{-iql}& 0 & J_y & -J_z & -\frac{J_x}{2}\;e^{-iql} \\
 0 & 0 & 0 & 0 &  -J_y & 0 & J_x & J_z  \\
0 & 0 & 0 & 0 & J_z & -J_x & 0 & J_y \\
 -\frac{J_x}{2}\;e^{iql} & 0 & 0 & 0 &  \frac{J_x}{2}\;e^{iql} & -J_z & -J_y & 0 \\
\end{pmatrix}
\label{Eqn:H_matrix_final}
\end{equation}
and $H_\uparrow$ is given by the same Hamltonian as above with all eight $q$ dependent terms set to zero.
  
The next step is to find the eigenvalues and eigenvectors of these matrices.  The eigenvalues are given by, 
\bea
& \lambda^\downarrow(q)=\pm \frac{1}{2}\sqrt{J^2_x+J^2_y+J^2_z \pm 2\cos(\frac{ql}{2})J_x\sqrt{J^2_y+J^2_z}},\quad 
 \pm\frac{1}{4}\left( J_x \pm \sqrt{J^2_x +4(J^2_y +J^2_z)}\right) \nonumber \\
  & \text{and} \quad \lambda^\uparrow(q)= \pm\frac{1}{4}\left( J_x \pm \sqrt{J^2_x +4(J^2_y +J^2_z)}\right), 
  \label{Eqn:eigenvalues}
 \eea
 where the $\lambda^\uparrow(q)$ values are doubly degenerate.
From the above equation, it is clear that there are twelve flat bands and four dispersive bands in this model. After
the imposition of the single occupancy constraint, it is only the four dispersive bands in the model which survive.
The simplest way to see this is to rewrite the fermions in terms of the Majorana fermions as shown in Appendix A of  Ref.\cite{bn}.
 Note that imposition of the diagonal $SU(2)$ constraint $n_{i\uparrow} +n_{i\downarrow}=1$ is equivalent to imposing
the Majorana constraint $b_i^xb_i^yb_i^zc_i =1$ on the wavefunctions (where the Majorana fermions $b_i^{x,y,z}$ and
$c_i$ have been defined in Appendix B).

\begin{figure}[!ht]
 \label{fig.gapless_tetra}
 \begin{center}
   \includegraphics[width=10cm]{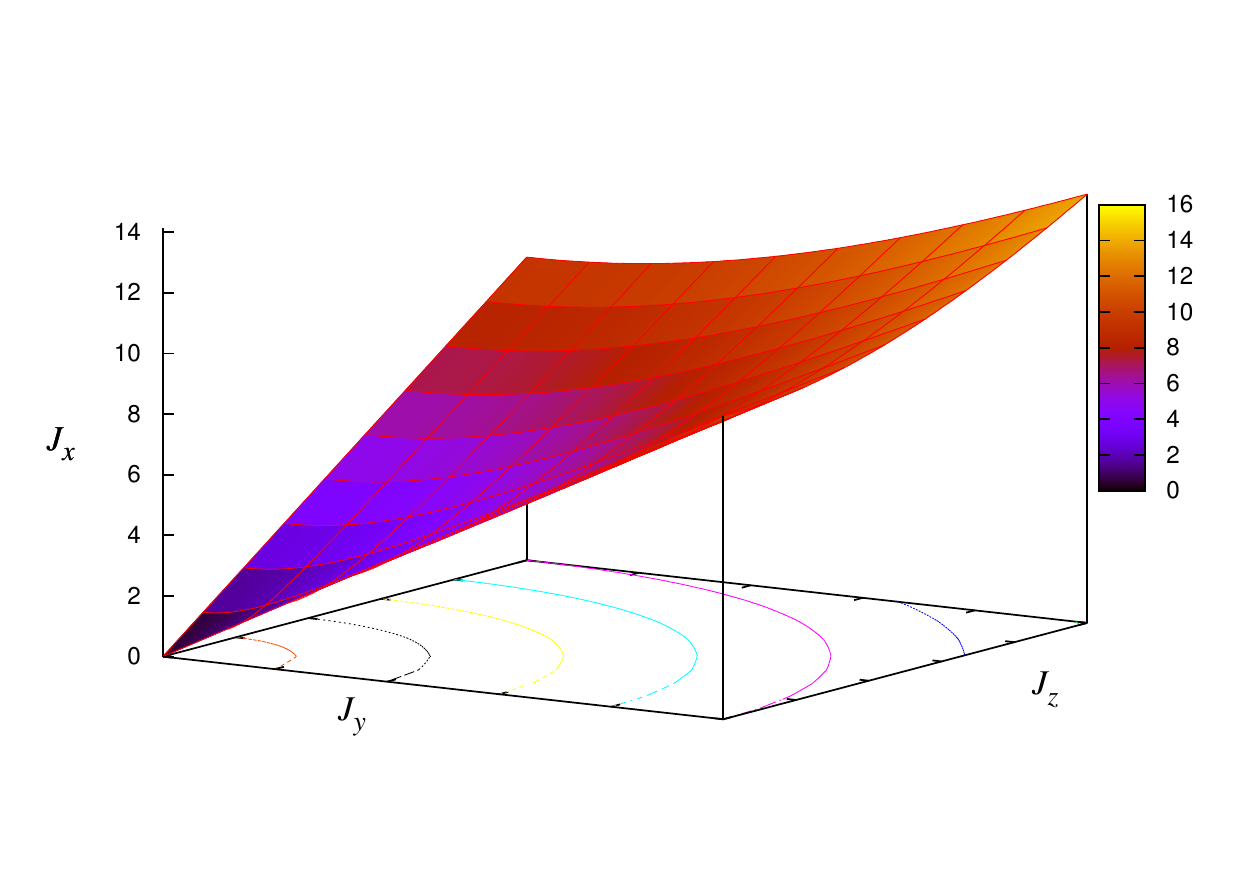}
   \end{center} 
   \caption{The gapless phase in tetrahdral model given by  $J_x=\sqrt{J^2_y+J^2_z}$ along the cone.} \label{gaplessphase}
  \end{figure}

 We will now concentrate only on the spin $\downarrow$ sector, which contains all the dispersive modes which survive the imposition
 of the single occupancy constraint.
  The gap in the spectrum is the difference between the smallest positive eigenvalue and the largest positive eigenvalue. From
  Eq. \ref{Eqn:eigenvalues}, one can obtain 
  \beq
  {\text{Gap}} =\lambda^{min}_{+}-\lambda^{max}_{-}=|\sqrt{J^2_y+J^2_z}-J_x|~.
  \eeq
   The
  gapless phase is thus identified by $J_x=\sqrt{J^2_y+J^2_z}$, which is plotted as a cone  in Fig.\ref{gaplessphase}. The two gapped 
  phases are on the either side of the cone. 
The eigenvalues in the gapless phase are given by
\begin{equation}
 \epsilon_{\text{gapless}}=0,\,0,\,\,\pm J_x,\,\,\pm\frac{\sqrt{3\pm\sqrt{5}}}{2 \sqrt{2}} J_x 
\end{equation}
Among these eight eigenvalues, $0,\,0,\,\,\pm J_x$ belong to the four dispersive bands and 
$\pm\frac{\sqrt{3\pm\sqrt{5}}}{2 \sqrt{2}}J_x$ belong to the four flat bands. The eigenvectors, 
$\left(\alpha \,\,\beta \,\,\gamma \,\,\delta \,\, \rho \,\, \sigma \,\, \theta \,\, \phi \,\, \right)^T$ corresponding to  these
eigenvalues can  also be  calculated and are given by
\begin{equation}
\epsilon=J_x, \quad  \text{Eigenvector:} \frac{1}{J_x}
\left(\begin{array}{c}
 -iJ_x \\
 -(J_y-i J_z) \\
 i (J_y-i J_z) \\
 J_x \\
 i (J_y-i J_z) \\
 -iJ_x \\
 -(J_y-i J_z) \\
 J_x\\
\end{array}\right), 
\epsilon=-J_x ,\quad \text{Eigenvector:}
\frac{1}{J_x}\left(\begin{array}{c}
 iJ_x \\
 -(J_y+i J_z) \\
 -i (J_y+iJ_z) \\
 J_x \\
 -i (J_y+i J_z) \\
 iJ_x \\
 -(J_y+i J_z) \\
 J_x\\
\end{array}\right) \nonumber
\end{equation}
\begin{equation}
 \epsilon=0,\quad \text{Eigenvector:}
 \frac{1}{J_y}\left(\begin{array}{c}
 J_y  \\
 0  \\
 J_x \\
- J_z  \\
 J_x \\
 J_y  \\
 0  \\
 -J_z \\
\end{array}\right), 
 \epsilon=0,\quad \text{Eigenvector:}
 \frac{1}{J_y}\left(\begin{array}{c}
  0  \\
 J_y  \\
 -J_z \\
J_x  \\
 -J_z\\
 0  \\
J_y  \\
J_x \\
\end{array}\right)\nonumber
\end{equation}

Now we note that the flat bands can be decoupled from the problem and we can rewrite the Hamiltonian by projecting
it onto the subspace involving only the four dispersing bands. This is the subspace of dispersing Majorana modes, our Hamiltonian is now 
indentical to the one studied in Ref.\cite{saket} by Saket {\it et al} and 
the phase diagram obtained here matches exactly with the phase diagram obtained earlier.

\section{Computation of the spin-spin correlation in the gapless phase}

We will now compute the spin-spin correlation in the gapless phase.  It is possible to project the Hamiltonian in terms of the
slave fermions to the single occupancy subspace and compute the spin-spin correlation
in terms of slave fermions.  However, it is easier to write the Hamiltonian of the four dispersing modes
in terms of Majorana fermions and then compute the spin-spin correlation.
In terms of the Majorana fermions, the Hamiltonian of the dispersing modes can be written as shown in Appendix B:
\begin{eqnarray}
 H=\sum_i i J_x \left(c_{i-1,4}c_{i,1}+c_{i,2}c_{i,3}\right)
 +i J_y\left(c_{i,1}c_{i,2}+c_{i,3}c_{i,4}\right) + i J_z(-c_{i,1}c_{i,3}+c_{i,2}c_{i,4}),
  \label{Eqn:Tetra_H_mj}
  \end{eqnarray}
  where $c_i$ are Majorana fermions. 
Fourier transforming the above $H$ gives,
\begin{eqnarray}
 H&=&i\sum_q J_x \left(e^{iql} c_{-q,4}c_{q,1}+c_{-q,2}c_{q,3}\right)
 + J_y\left(c_{-q,1}c_{q,2}+c_{-q,3}c_{q,4}\right) +  J_z(-c_{-q,1}c_{q,3}+c_{-q,2}c_{q,4})
   \label{Eqn:Tetra_H_mjq}
  \end{eqnarray}

We first note that since the $Z_2$ fluxes on each plaquette are conserved quantities, the spin-spin correlation function which
is actually an overlap of two states on sites $i$ and $j$ is zero unless the two sites happen to be nearest neighbours~\cite{mandal}.
In other words, there is no correlation at all beyond nearest neighbours.  So we can only compute the dynamical  spin-spin correlation
on nearest neighbour sites, which turns out to be of the form
\beq
\langle S^\alpha_i S^\alpha_j \rangle = g_{\langle i,j \rangle}(t) \delta_{\alpha\beta}~.
\eeq
Even without any further computation, we can see that the gapless sector describes a spin liquid, since there is no long
range correlation beyond nearest neighbour correlation.

The eigenvalues of the Hamiltonian  was calculated in the  earlier section. The eigenvectors can also be found, now in terms of the Majorana fermions  and
the Hamiltonian can be written as  $H=\sum_n \chi^\dagger_n H_{nm} \chi_m$.
In the gapless phase the eigenvalues are, $0,0,\pm J_x$ and the eigenvector matrix is given by 
\begin{eqnarray}\frac{1}{2J_x}\begin{pmatrix}
 iJ_x & 0 & J_y\sqrt{2}& -i J_x\\
 -(J_y+i J_z) & \sqrt{2}J_y & 0 & -J_y+i J_z \\
 -i J_y+J_z & -J_z\sqrt{2} & J_x\sqrt{2} & i J_y+J_z \\
 J_x & J_x\sqrt{2} & -J_z\sqrt{2} & J_x \\
 \end{pmatrix}
 \end{eqnarray}
 In order to calculate the spin spin correlator $\langle S^z_i S^z_j \rangle$ we now  go to a diagonal basis so that the Hamiltonian has 
 the form $H=\sum_n \epsilon_n \psi^\dagger_n \psi_n$. The eigenvectors in the new basis are found to be 
\beq
\begin{pmatrix}
\psi_1\\
\psi_2\\
\psi_3\\
\psi_4\\ 
\end{pmatrix}
=
\frac{1}{2J_x}
 \begin{pmatrix}
\sqrt{2} J_y c_{q,3} +i J_x (c_{q,1}- c_{q,4}) \\
 J_y\left(c_{q,2} \sqrt{2}-c_{q,1}-c_{q,4} \right)-iJ_z\left(c_{q,1}-c_{q,4} \right) \\
 J_x\,\sqrt{2}\,c_{q,3}+J_z \left( c_{q,1}+c_{q,4}-c_{q,2}\,\sqrt{2} \right)
 -iJ_y\left(c_{q,1}-c_{q,4}\right) \\
J_x\left(c_{q,1}+c_{q,4}+ c_{q,2} \,\sqrt{2}\right)- J_z\,\sqrt{2}\,c_{q,3}  \\
\end{pmatrix}
\eeq
 The partition function $Z = Tr e^{-H}$ 
over the $2^4=16$ four fermionic states $|0000\rangle, |1000\rangle, |0100\rangle \cdots |1111\rangle$ can now be found to be $Z=6+8\cosh{J_x}$. 
The spin-spin correlators  are hence found to be (for nearest neighbours)
\bea
\langle S^z_i S^z_j \rangle=\frac{d}{d J_z} \ln Z
&=&\frac{8\sinh{J_x}}{6+8\cosh{J_x}}\frac{J_z}{\sqrt{J^2_y +J^2_z}} \nonumber \\
\langle S^y_i S^y_j \rangle=\frac{d}{d J_z} \ln Z
&=&\frac{8\sinh{J_x}}{6+8\cosh{J_x}}\frac{J_y}{\sqrt{J^2_y +J^2_z}} \nonumber \\
\langle S^x_i S^x_j \rangle=\frac{d}{d J_z} \ln Z
&=&\frac{8\sinh{J_x}}{6+8\cosh{J_x}}~.
\eea
 Since our gapless phase is just a circle, these are just numbers for each value of $J_x,J_y$ and $J_z$ and have no further structure.
 They can be dynamical correlation functions if the $J_x,J_y,J_z$ are time-dependent.
 But our main conclusion is that our system is a spin liquid in the sense that there is no long-range magnetic order and the spin-spin correlations are non-zero
 only for nearest neighbours.

\section{Mapping to the  one-dimensional Kitaev model}
In this section we analyze the similarities between the tetrahedral chain 
and the 1D Kitaev $p$-wave superconductor~\cite{onedkitaev}. The tetrahedral spin chain is similar to the 1D Kitaev model,
except that alternate bonds are `dressed\rq{} by the plaquette operators.
The 1D Kitaev model consists of spinless fermions connected in a one-dimensional  chain. Hopping and p-wave pairing
are allowed between the nearest neighbour sites. The Hamiltonian takes the following form,
\begin{equation}
 H=-\sum_j \left[(t f^\dagger_j f_{j+1} + t^* f^\dagger_{j+1} f_{j})-(\Delta f^\dagger_j f^\dagger_{j+1}+
 \Delta^* f_{j+1} f_{j})+\mu(f^\dagger_j f_j -\frac{1}{2})  \right] 
 \label{Eqn:Kitaev_1DH_c}
\end{equation}
  where $t$ is the hopping amplitude, $\Delta$ is the p-wave pairing and $\mu$, the chemical potential. This is structurally similar to 
either $H_\uparrow$ or $H_\downarrow$ of the tetrahedral model, where, for instance $H_\downarrow $ is given by 
 \begin{eqnarray}
 H_\downarrow &=&-\sum_i t^x_\downarrow \;(f^{\dagger}_{i-1,4\;\downarrow}f_{i,1\;\downarrow}
 +f^{\dagger}_{i,2\;\downarrow}f_{i,3\;\downarrow})
 +t^{x^\ast}_\downarrow\; (f^{\dagger}_{i,1\;\downarrow}f_{i-1,4\;\downarrow}+f^{\dagger}_{i,3\;\downarrow}f_{i,2\;\downarrow})
 +\Delta^x_\downarrow \;f^{\dagger}_{i-1,4\;\downarrow}f^{\dagger}_{i,1\;\downarrow}\nonumber\\
&& + \Delta^{x^\ast}_\downarrow f_{i,1\;\downarrow}f_{i-1,4\;\downarrow}
+ t^y_\downarrow \;(f^{\dagger}_{i,1\;\downarrow}f_{i,2\;\downarrow}+f^{\dagger}_{i,3\;\downarrow}f_{i,4\;\downarrow})
+t^{y^\ast}_\downarrow\; (f^{\dagger}_{i,2\;\downarrow}f_{i,1\;\downarrow}+f^{\dagger}_{i,4\;\downarrow}f_{i,3\;\downarrow})\nonumber \\
 &&+ t^z_\downarrow \;(f^{\dagger}_{i,1\;\downarrow}f_{i,3\;\downarrow}+f^{\dagger}_{i,2\;\downarrow}f_{i,4\;\downarrow})
+t^{z^\ast}_\downarrow\; (f^{\dagger}_{i,3\;\downarrow}f_{i,1\;\downarrow}+f^{\dagger}_{i,4\;\downarrow}f_{i,2\;\downarrow})\nonumber \
\end{eqnarray}
which looks similar to the Kitaev model, except that the chemical potential term is not really an ordinary  chemical potential term, but instead
involves hoppings within the plaquette $i$. Hence, the Hamiltonian can be rewritten as 
\beq
H_\downarrow 
= t^x_\downarrow \;f^{\dagger}_{i-1,4\;\downarrow}f_{i,1\;\downarrow}+t^{x^\ast}_\downarrow\;f^{\dagger}_{i,1\;\downarrow}f_{i-1,4\;\downarrow}
+\Delta^x_\downarrow \;f^{\dagger}_{i-1,4\;\downarrow}f^{\dagger}_{i,1\;\downarrow}+
\Delta^{x^\ast}_\downarrow f_{i,1\;\downarrow}f_{i-1,4\;\downarrow}+\mu f^{\dagger}_{i,\alpha \;\downarrow}f_{i,\beta\;\downarrow}
\label{Eqn:1dchainH}
\eeq
with 
  \begin{eqnarray}
  \label{Eqn:1dchainmu}
 \mu\, f^{\dagger}_{i,\alpha \;\downarrow}f_{i,\beta\;\downarrow}&=&t^x_\downarrow \;f^{\dagger}_{i,2\;\downarrow}f_{i,3\;\downarrow}
 +t^{x^\ast}_\downarrow\; f^{\dagger}_{i,3\;\downarrow}f_{i,2\;\downarrow}
 + t^y_\downarrow \;(f^{\dagger}_{i,1\;\downarrow}f_{i,2\;\downarrow}+f^{\dagger}_{i,3\;\downarrow}f_{i,4\;\downarrow})
+t^{y^\ast}_\downarrow\; (f^{\dagger}_{i,2\;\downarrow}f_{i,1\;\downarrow}+f^{\dagger}_{i,4\;\downarrow}f_{i,3\;\downarrow})\nonumber\\
&&+ t^z_\downarrow \;(f^{\dagger}_{i,1\;\downarrow}f_{i,3\;\downarrow}+f^{\dagger}_{i,2\;\downarrow}f_{i,4\;\downarrow})
+t^{z^\ast}_\downarrow\; (f^{\dagger}_{i,3\;\downarrow}f_{i,1\;\downarrow}+f^{\dagger}_{i,4\;\downarrow}f_{i,2\;\downarrow})  
  \end{eqnarray}
Therefore the tetrahedral chain can be mapped into a special case of 1D Kitaev p-wave superconductor.  
In Kitaev's original model, all the parameters, $i.e.$,  $t$, $\Delta$ and $\mu$  are real and the 
fermion operators are spinless. Both these conditions are violated in this model derived from the tetrahedral chain;  here, 
there are two spin channels ($H_{\uparrow,\downarrow}$) which are independently mapped into two Kitaev models,
and the parameters $t^{x,y,z}$ and $\Delta^{x}$ in each one of them are fully complex.  Note that this mapping is from the full fermion model and not the spin model, because the single occupancy constraint has not been imposed.
The generalisation of one-dimensional Kitaev models subjected to different potentials has been addressed by
deGottardi {\it et al}~\cite{degottardi2013} who specifically  studied the Kitaev p-wave chain with complex parameters in the
chemical  potential, as an example of the Kitaev model  with time-reversal symmetry breaking (belonging to class {\bf D}).
As shown by them,  the p-wave pairing $\Delta$ and $\Delta^*$ 
  can be made real by the global phase transformation $f^\dagger \rightarrow f^\dagger e^{-i\theta}$ and $f \rightarrow f e^{i\theta}$. The 
 complex hopping terms, however, are invariant under this transformation and can 
 be written as,  $t=t_0 e^{i\phi}$, where $t_0$ is a real positive number. In 
  Ref.\onlinecite{degottardi2013}, a detailed study of the  phase diagram of Kitaev model for the full range of values of $\phi$ has been made.
  We find that our model  
 where $t_{\uparrow,\downarrow}$ are fully imaginary, maps to the case where $\phi=\frac{\pi}{2}$. The phase diagram 
 of the  original  Kitaev model (where $\phi=0$  with real $t$ and $\Delta$) consists of both trivial and topological phases. 
 As the value  of $\phi$ increases to $\frac{\pi}{2}$ (where $t$ and $\Delta$ are fully complex), the width of the topological phase decreases 
 and finally becomes zero. Therefore our model falls into the trivial part of the phase diagram, with no topology and hence, no Majorana modes
 at the edges.

\section{Discussions and conclusion}

In this paper, we have studied a slave fermion approach to the tetrahedral spin chain and obtained its phase diagram and
showed that it matches the earlier results obtained using the Majorana fermions. We have then computed the spin-spin
correlation function in the gapless phase, which falls of beyond nearest neighbours, and hence confirms that the
gapless phase is a spin liquid. We have also mapped our model to the standard
one-dimensional Kitaev model, so that we can read off the topological nature of the model. We find that the ground state of the model
falls in the topologically trivial part of the standard Kitaev model, and hence our model does not have edge Majorana modes.

\section*{Acknowledgments} We would like to thank R. Shankar for useful discussions and for sending us his notes. We would also like
to thank S. Mandal for helpful correspondence. PM would like to thank A. Joshi for all discussions and suggestions.


 \appendix

\section{Spin operators and spin Hamiltonian in terms of slave fermions}

In this appendix, we give some details of the representation of the spin operators in terms of slave fermions and
show how the mean field Hamiltonian in Eq.\ref{Eqn:H_matrix} in the main text is obtained.

Our
 starting point is to use the slave fermion formalism where the spins are written in terms of standard fermions, rather than Majorana fermions, $i.e.$,
 we rewrite the quadratic terms of the Hamiltonian in Eq.\ref{Eqn:tetra_H}, the
 $\sigma^x_i \sigma^x_j$, $\sigma^y_i \sigma^y_j$ and $\sigma^z_i \sigma^z_j$,  on the $x,y$ and $z$ links respectively,
 in terms of standard  fermion operators as
 \begin{eqnarray}
 \sigma^x_i \sigma^x_j &=& -\frac{1}{4}[f^{\dagger}_{i\uparrow}f^{\dagger}_{j\uparrow}f_{i\downarrow}f_{j\downarrow}+
 f^{\dagger}_{i\downarrow}f^{\dagger}_{j\downarrow}f_{i\uparrow}f_{j\uparrow}+
 f^{\dagger}_{i\uparrow}f_{j\uparrow}f^{\dagger}_{j\downarrow}f_{i\downarrow}+
 f^{\dagger}_{i\downarrow}f_{j\downarrow}f^{\dagger}_{j\uparrow}f_{i\uparrow}]\nonumber\\
 \sigma^y_i \sigma^y_j &=&-\frac{1}{4}[-f^{\dagger}_{i\uparrow}f^{\dagger}_{j\uparrow}f_{i\downarrow}f_{j\downarrow}-
 f^{\dagger}_{i\downarrow}f^{\dagger}_{j\downarrow}f_{i\uparrow}f_{j\uparrow}+
 f^{\dagger}_{i\uparrow}f_{j\uparrow}f^{\dagger}_{j\downarrow}f_{i\downarrow}+
 f^{\dagger}_{i\downarrow}f_{j\downarrow}f^{\dagger}_{j\uparrow}f_{i\uparrow}]\nonumber\\
 \sigma^z_i \sigma^z_j &=&-\frac{1}{4}[-f^{\dagger}_{i\uparrow}f^{\dagger}_{j\uparrow}f_{j\downarrow}f_{i\downarrow}-
 f^{\dagger}_{i\downarrow}f^{\dagger}_{j\downarrow}f_{j\uparrow}f_{i\uparrow}+
 f^{\dagger}_{i\uparrow}f_{j\uparrow}f^{\dagger}_{j\uparrow}f_{i\uparrow}+
 f^{\dagger}_{i\downarrow}f_{j\downarrow}f^{\dagger}_{j\downarrow}f_{i\downarrow}]
 \label{Eqn:sigma_xx_yy_zz}
 \end{eqnarray}
 using $\hat{\sigma}^\kappa_i=\frac{1}{2}f^{\dagger}_{i\alpha}
 \sigma^\kappa_{\alpha \beta} f_{i\beta}$. where the 
 operators $f_{i\alpha}$ and $f^{\dagger}_{i\alpha}$ are the fermion annihilation and creation operators. The index $i$ denotes
 the site, indices $\kappa=(x,y,z)$  denote the three components of the spin and the indices $\alpha,\,\beta = \uparrow,\downarrow$ denote
 the spin projections.
We now perform a mean field analysis of the Hamiltonian by introducing the Hubbard-Stratanovich ($\mathcal{HS}$)  fields in the 
hopping and the $p$-wave superconducting channels. As explained in Ref.[\onlinecite{bn}], in the
absence of spin-rotational symmetry, independent $\mathcal{HS}$ fields have to be introduced for the fermion operators
on the $x,y$ and $z$ links. The $\mathcal{HS}$ transformed form of the spin operators on the $x$ links is given by
 \begin{eqnarray}
 J_x\; \sigma^x_i \sigma^x_j &=&-\frac{8}{J_x}(|\phi_1|^2+|\phi_2|^2)
   +\phi_1(f^{\dagger}_{i\uparrow}f_{j\uparrow}+f^{\dagger}_{i\downarrow}f_{j\downarrow})
   +i\phi_2(f^{\dagger}_{i\uparrow}f_{j\uparrow}-f^{\dagger}_{i\downarrow}f_{j\downarrow})
   +\phi^{\dagger}_1(f^{\dagger}_{j\uparrow}f_{i\uparrow}+f^{\dagger}_{j\downarrow}f_{i\downarrow})
  +i\phi^{\dagger}_2(f^{\dagger}_{j\uparrow}f_{i\uparrow}-f^{\dagger}_{j\downarrow}f_{i\downarrow})\nonumber\\
  &&-\frac{8}{J_x}(|\theta_1|^2+|\theta_2|^2)+
  \theta_1(f^{\dagger}_{i\uparrow}f^{\dagger}_{j\uparrow}+f^{\dagger}_{i\downarrow}f^{\dagger}_{j\downarrow})
   +i\theta_2(f^{\dagger}_{i\uparrow}f^{\dagger}_{j\uparrow}-f^{\dagger}_{i\downarrow}f^{\dagger}_{j\downarrow})
   +\theta^{\dagger}_1(f_{j\uparrow}f_{i\uparrow}+f_{j\downarrow}f_{i\downarrow})
  +i\theta^{\dagger}_2(f_{j\uparrow}f_{i\uparrow}-f_{j\downarrow}f_{i\downarrow})\nonumber\\
 &=&-\frac{8}{J_x}(|\phi_1|^2+|\phi_2|^2)+(\phi_1+i\phi_2)f^{\dagger}_{i\uparrow}f_{j\uparrow}+
 (\phi_1-i\phi_2)f^{\dagger}_{i\downarrow}f_{j\downarrow}+(\phi^{\dagger}_1+i\phi^{\dagger}_2)f^{\dagger}_{j\uparrow}f_{i\uparrow}+
 (\phi^{\dagger}_1-i\phi^{\dagger}_2)f^{\dagger}_{j\downarrow}f_{i\downarrow}\nonumber\\
 &&-\frac{8}{J_x}(|\theta_1|^2+|\theta_2|^2)
 +(\theta_1+i\theta_2)f^{\dagger}_{i\uparrow}f^{\dagger}_{j\uparrow}+(\theta_1-i\theta_2)f^{\dagger}_{i\downarrow}f^{\dagger}_{j\downarrow}+
 (\theta^{\dagger}_1+i\theta^{\dagger}_2)f_{j\uparrow}f_{i\uparrow}
 +(\theta^{\dagger}_1-i\theta^{\dagger}_2)f_{j\downarrow}f_{i\downarrow}, 
 \label{Eqn:Jx_HS_0}
  \end{eqnarray}
where the $\phi_i$ and the $\theta_i$ fields are the bosonic $\mathcal{HS}$ fields evaluated on the $x$ links in the hopping
and superconducting channels respectively. We can check that this decoupling gives back the original term by integrating out
the bosonic fields.

\section{Majoranization of Spins}
\label{app:majo_spin}

In this Appendix,  we briefly review  how  Majorana fermionization  can be used to solve the spin model. 
We start with the spin Hamiltonian in Eq.\ref{Eqn:tetra_H}, which is given below:
  \begin{equation}
 H=\sum_i -J_x(\sigma^x_{i-1,4}\;\sigma^x_{i,1}+\sigma^x_{i,2}\;\sigma^x_{i,3})
 -J_y(\sigma^y_{i,1}\;\sigma^y_{i,2}+\sigma^y_{i,3}\;\sigma^y_{i,4})
 -J_z(\sigma^z_{i,1}\;\sigma^z_{i,3}+\sigma^z_{i,2}\;\sigma^z_{i,4}).
  \label{Eqn:app_Tetra_H}
 \end{equation}
 The spin operators in the above Hamiltonian can directly be  expressed in terms of Majorana fermions. The spin operators in terms of Majorana fermions
 have the following forms:
 \begin{equation}
  \sigma^x=i b^x c, \quad \sigma^y=i b^y c, \quad \sigma^z=i b^z c
  \label{Eqn:app_spin_majo}
 \end{equation}
Substituting this back in Eq.\ref{Eqn:app_Tetra_H} gives
\begin{equation}
 H=\sum_i -J_x(b^x_{i-1,4}b^x_{i,1}\;c_{i-1,4}c_{i,1}+b^x_{i,2}b^x_{i,3}\;c_{i,2}c_{i,3})
 -J_y(b^y_{i,1}b^y_{i,2}\;c_{i,1}c_{i,2}+b^y_{i,3}b^y_{i,4}\;c_{i,3}c_{i,4})
 -J_z(b^z_{i,1}b^z_{i,3}\;c_{i,1}c_{i,3}+b^z_{i,2}b^z_{i,4}\;c_{i,2}c_{i,4}).
  \label{Eqn:app_Tetra_H_mj}
 \end{equation}
where one has four fermion operators. Now, there are two ways to proceed which are given below:
  
\subsection*{Method 1: Kitaev's method}
  \label{app:majo_spin_1}
  Let us  rename  $i b^x_{i,\alpha}b^x_{j,\beta}=u^x{(i_\alpha,j_\beta)}, \quad ib^y_{i,\alpha}b^y_{j,\beta}=u^y{(i_\alpha,j_\beta)}, \quad 
i b^z_{i,\alpha}b^z_{j,\beta}=u^z{(i_\alpha,j_\beta)}$. In terms of the bilinear operators $u^{x,y,z}(i_\alpha,j_\beta)$, 
 Eq.\ref{Eqn:app_Tetra_H_mj} can be rewritten as 
 \begin{eqnarray}
 H&=&\sum_i i J_x \left(u^x{({i-1}_4,i_1)}\;c_{i-1,4}c_{i,1}+u^x{(i_2,i_3)}\;c_{i,2}c_{i,3}\right)
 +i J_y\left(u^y{(i_1,i_2)}\;c_{i,1}c_{i,2}+u^y{(i_3,i_4)}\;c_{i,3}c_{i,4}\right) \nonumber\\
 &+& i J_z(u^z{(i_1,i_3)}\;c_{i,1}c_{i,3}+u^z{(i_2,i_4)}\;c_{i,2}c_{i,4}).
  \label{Eqn:app_Tetra_H_mj1}
 \end{eqnarray}
 The operators $u^{x,y,z}{(i_\alpha,j_\beta)}$ for a given set of indices $i$, $j$, $\alpha$, $\beta$ appear only once in the Hamiltonian.
 Therefore the operators $u^{x,y,z}{(i_\alpha,j_\beta)}$ commute with the Hamiltonian and  share the same set of eigenvectors.
If $H|\psi\rangle=\varepsilon|\psi\rangle$, the effect of $u^{\tau}{(i_\alpha,j_\beta)}$ on $|\psi\rangle$ is calculated in the following
 way:
 \begin{eqnarray}
 \label{Eqn:app_U}
  u^{\tau^2}{(i_\alpha,j_\beta)} |\psi\rangle &=& u^{\tau}{(i_\alpha,j_\beta)}\;u^{\tau}{(i_\alpha,j_\beta)}|\psi\rangle \nonumber\\
  &=& i\; b^\tau_{i,\alpha}b^\tau_{j,\beta} \;i\; b^\tau_{i,\alpha}b^\tau_{j,\beta}|\psi\rangle \nonumber\\
   &=&  b^\tau_{i,\alpha} b^\tau_{i,\alpha}\;b^\tau_{j,\beta} b^\tau_{j,\beta}|\psi\rangle= 1 |\psi\rangle
 \end{eqnarray}
Therefore $u^{\tau}{(i_\alpha,j_\beta)}|\psi\rangle=\pm 1|\psi\rangle$. The ground state is translationally invariant and we can choose
$u^{\tau}{(i_\alpha,j_\beta)}=1$ for all values of $i,j,\alpha,\beta$.
This simplifies the Hamiltonian in  Eq.\ref{Eqn:app_Tetra_H_mj1} to the quadratic form given as 
\begin{eqnarray}
 H=\sum_i i J_x \left(c_{i-1,4}c_{i,1}+c_{i,2}c_{i,3}\right)
 +i J_y\left(c_{i,1}c_{i,2}+c_{i,3}c_{i,4}\right) + i J_z(c_{i,1}c_{i,3}+c_{i,2}c_{i,4}).
  \label{Eqn:app_Tetra_H_mj2}
  \end{eqnarray}
 This  quadratic Hamiltonian  can be diagonalized and solved to find the dispersion relation.

\subsection*{Method 2: H-S transformation}
   \label{app:majo_spin_2}
    
 To use Kitaev\rq{}s  method discussed above, one needs to understand the symmetries of the model. Futhermore, from the 
 resulting $H$ in Eq.\ref{Eqn:app_Tetra_H_mj2}, it is not possible to make a connection with the Hamiltonian in 
 Eq.\ref{Eqn:H_matrix}. In this section, we take a midway path between Kitaev's method and the slave-fermion method 
 used in the main text. 
  
  Here, we  start with the Majoranized 
  Hamiltonian in Eq.\ref{Eqn:app_Tetra_H_mj} and then  decouple it, using  the $\mathcal{HS}$ transformation, 
 in the mean field theory limit. 
This would be another way of reaching the quadratic limit in the main paper given 
  in   Eq.\ref{Eqn:H_matrix} (after a linear transformation).
     More explicitly, let us start with applying the $\mathcal{HS}$  transformation on Eq.\ref{Eqn:app_Tetra_H_mj} to decouple the 
four fermion terms. This leads to 
   \begin{eqnarray}
\label{Eqn:app_sxsx}
J_x \; b^x_{i,\alpha}b^x_{j,\beta}\;c_{i,\alpha}c_{j,\beta} &=&
-\frac{\tilde{\Phi}^*_1 {\Phi}_1+{\Phi}^*_1\tilde{\Phi}_1 }{2J_x}+\frac{\Phi_1 \; b^x_{i,\alpha} b^x_{j,\beta}}{2}
-\frac{\Phi^*_1 \; b^x_{j,\beta}b^x_{i,\alpha} }{2}
+\frac{\tilde{\Phi}_1 \; c_{i,\alpha}c_{j,\beta}}{2}-\frac{\tilde{\Phi}^*_1 \;  c_{j,\beta}c_{i,\alpha} }{2} \\
\label{Eqn:app_sysy}
J_y \; b^y_{i,\alpha}b^y_{j,\beta}\;c_{i,\alpha}c_{j,\beta} &=&
-\frac{\tilde{\Phi}^*_2 {\Phi}_2+{\Phi}^*_2\tilde{\Phi}_2}{2J_y}+\frac{\Phi_2 \; b^y_{i,\alpha} b^y_{j,\beta}}{2}
-\frac{\Phi^*_2 \; b^y_{j,\beta}b^y_{i,\alpha} }{2}
+\frac{\tilde{\Phi}_2 \; c_{i,\alpha}c_{j,\beta}}{2}-\frac{\tilde{\Phi}^*_2 \;c_{j,\beta} c_{i,\alpha}  }{2} \\
\label{Eqn:app_szsz} {\rm and}
J_z \; b^z_{i,\alpha}b^z_{j,\beta}\;c_{i,\alpha}c_{j,\beta} &=&
-\frac{\tilde{\Phi}^*_3 {\Phi}_3+{\Phi}^*_3\tilde{\Phi}_3}{2J_z}+\frac{\Phi_3 \; b^z_{i,\alpha} b^z_{j,\beta}}{2}
-\frac{\Phi^*_3 \;b^z_{j,\beta} b^z_{i,\alpha} }{2}
+\frac{\tilde{\Phi}_3 \; c_{i,\alpha}c_{j,\beta}}{2}-\frac{\tilde{\Phi}^*_3 \;  c_{j,\beta} c_{i,\alpha} }{2},
    \end{eqnarray}
 where, in  the mean field limit, $\Phi_\alpha$ and $\tilde{\Phi}_\alpha$ are replaced by  $\langle\Phi_\alpha\rangle$ and $\langle \tilde{\Phi}_\alpha \rangle$ respectively. This Hamiltonian is quadratic, but  in  terms of Majorana fermions. 
    To compare this Hamiltonian with that of in Eq.\ref{Eqn:H_matrix}, we need to rewrite the  Majorana fermions in terms
of Dirac fermions. 
     The four Majorana fermions, with momenta $q$ and sublattice indices $i$ ($b^x_{qi}$, $b^y_{qi}$, $b^z_{qi}$ and $c_{qi}$) can  be
represented as  linear combination of Dirac fermions in the following way:
  \begin{equation}
   b^x_{qi}=i(f^{\dagger}_{q,i\;\uparrow}-f_{-q,i\;\uparrow}),\quad b^y_{qi}=(f^{\dagger}_{q,i\;\uparrow}+f_{-q,i\;\uparrow}),\quad 
   b^z_{qi}=(f^{\dagger}_{q,i\;\downarrow}+f_{-q,i\;\downarrow}),\quad c_{qi}=i(f^{\dagger}_{q,i\;\downarrow}-f_{-q,i\;\downarrow})~.
   \label{Eqn:majorana_form}
   \end{equation}
Substituting the above expressions for the Majorana fermions in Eqs.\ref{Eqn:app_sxsx}-\ref{Eqn:app_szsz} will
 give the Hamiltonian in terms of $f^{\dagger}_{\uparrow\;\downarrow}$ and $f^{\dagger}_{\uparrow\;\downarrow}$. 
However the resultant Hamiltonian
   will not be identical to  Eq.\ref{Eqn:H_matrix}  in the main text.  That is not surprising because the  Majorana representation of spins 
(in Eq.\ref{Eqn:app_spin_majo})  is valid only in the singly occupied subspace.  Hence, no further projection to the physical subspace is required.
Eq.\ref{Eqn:H_matrix} in the main text, however, still requires projection to the singly occupied subspace in order to 
represent the spin model. In other words, by rewriting the Majorana fermions after $\mathcal{HS}$ transformation in terms of ordinary fermions,
we get the quadratic fermion model which is already in the singly occupied subspace, with no further gauge freedom.

 A saddle point evaluation of the Eqs.\ref{Eqn:app_sxsx}-\ref{Eqn:app_szsz} give the following:
\begin{eqnarray}
 \langle\tilde{\Phi}_1 \rangle = J_x \langle b^x_{i,\alpha} b^x_{j,\beta} \rangle,&&\qquad\langle {\Phi}_1 \rangle= J_x \langle c_{i,\alpha} c_{j,\beta} \rangle \nonumber \\
\langle\tilde{\Phi}_2 \rangle= J_y \langle b^y_{i,\alpha} b^y_{j,\beta} \rangle,&&\qquad  \langle{\Phi}_2 \rangle= J_y \langle c_{i,\alpha} c_{j,\beta} \rangle \nonumber \\
\langle\tilde{\Phi}_3 \rangle= J_z \langle b^z_{i,\alpha} b^z_{j,\beta} \rangle,&&\qquad \langle{\Phi}_3 \rangle= J_z \langle c_{i,\alpha} c_{j,\beta} \rangle
\label{Eqn:app_majo_sp}
\end{eqnarray}
The equations in Eq.\ref{Eqn:app_majo_sp} can be solved self-consistently, even if the symmetries of the model
are not explicitly used. We mention this method as an alternative way of using the Majorana representation of the spins.


\begin{thebibliography}{99}

\bibitem{reviews}  F. Mila, Eur. Jnl. Phys. {\bf 21}, 499 (2000);   L Balents, Nature  {\bf 464}, 199 (2010); F. Misguich, cond-mat/0809.2257; P. A. Lee, Jnl of Phys. :
Conf. Series {\bf 529}, 012001 (2014).

\bibitem{gauge} J. B. Kogut, Rev. Mod. Phys {\bf 51}, 659 (1979); For more recent developments, see, for instance, 
 X. G. Wen, Quantum Field theory of Many Body Systems ( Oxford University Press).

\bibitem{sb} X. G. Wen, F. Wilczek and A. Zee, \prb{\bf 39}, 11413 (1989); 
G. Baskaran, Z. Zou and P. W. Anderson, Solid State Comm. {\bf 63}, 973 (1987); I. Affleck and J. B. Marston, \prb{\bf 37}, 3774 (1998); E. Dagotto, E. Fradkin and A. Moreo, \prb{\bf 38}, 2926 (1988).

\bibitem{sf} D. P. Arovas and   A. Auerbach, \prb{\bf 38} 316;
N. Read and S. Sachdev, \prl{\bf 62}, 1694 (1989); N. Read and S. Sachdev, \prl{\bf 66}, 1773 (1991).

\bibitem{qd} S. A. Kivelson, D. S. Rokhsar, and J. P. Sethna, \prb{\bf 35}, 8865 (1987); R. Moessner and S. L. Sondhi, \prl{\bf 86}, 1881 (2001).

\bibitem{numerics} G. J. Chen {\it et al}, \prb{\bf 42}, 2662 (1990); G. Misguich {\it et al}, \prb{\bf 60}, 1064 (1999); F. Becca {\it et al}, \prb 62, 15277 (2000); T. Kashima and M. Imada, J. Phys. Soc. Jpn. {\bf 70}, 3398 (2001).

\bibitem{2dkitaev} A. Kitaev, Ann. Phys. (N. Y.) {\bf 303}, 2 (2003); ibid, {\bf 321}, 2 (2006).

\bibitem{qc} For reviews, see C. Nayak, S. H. Simon, A. Stern, M. Freedman, S. D. Sarma, Rev. Mod. Phys. {\bf 80}, 1083 (2008);  J. K. Pachos,  Introduction to topological quantum computation, Cambridge University Press, 2012;  J. Alicea, arXiv:cond-mat/1202.1293 (2012).

\bibitem{bn} F. J. Burnell and C. Nayak, \prb{\bf 84}, 125125 (2011).

\bibitem{mandal} S. Mandal, R. Shankar and G. Baskaran, J. Phys. A:Math. Theor. {\bf 45} 335304 (2012).

\bibitem{genKitaev} H. Yao and S. Kivelson, \prl{\bf 99}, 247203 (2007); S. Yang, D. L. Zhou and C. P. Sun, \prb {\bf 76}, 180404 (R) (2007); S. Mandal and N. Surendran, \prb{\bf 79} 024426 (2009); Z. Nussinov and G. Ortiz, \prb{\bf 79}, 224408 (2009); G. Baskaran, G. Santhosh and R. Shankar, arXiv:0908.1614 (2009).

\bibitem{saket1} A.  Saket, S. R. Hassan and R. Shankar, \prb{\bf 82}, 174409 (2010).

\bibitem{saket2} A.  Saket, S. R. Hassan and R. Shankar, \prb{\bf 87}, 174414 (2013).

\bibitem{coldatoms} L. M. Duan, E. Demler and M. D. Lukin, \prl{\bf 91}, 090402 (2003).

\bibitem{qcircuits} J. Q. You, X. Shi, X. Hu and F. Nori, \prb{\bf 79}, 224408 (2009).



\bibitem{onedmodels} For a recent review of one dimensional magnetism, see H. J. Mikeska and A. K. Kolezhuk, chapter
on quantum magnetism, (1-83), Vol. 645 of Lecture notes in Physics, Springer (2004).

\bibitem{onedkitaev} A. Kitaev, Phys. Usp. {\bf 44}, 131 (2001).


\bibitem{saket} A. Saket, PhD thesis, Homi Bhabha National Institute, India,  April 2013.

\bibitem{degottardi2013} W. DeGottardi, M. Thakurathi, S. Vishveshwara, and D. Sen, \prb{\bf 88}, 165111 (2013).

\end{thebibliography}
\end{document}